\documentclass[aps, superscriptaddress,nofootinbib, reprint]{revtex4-2}
\usepackage{graphicx}
\usepackage{xcolor}
\usepackage[%
colorlinks=true,
linkcolor=blue,
citecolor=blue,
]{hyperref}
\usepackage{booktabs}
\usepackage{rotating}
\usepackage[marginal]{footmisc}
\usepackage{graphicx}
\usepackage{dcolumn}
\usepackage{bm}
\usepackage{xcolor} 
\newcommand{\rev}[1]{\textcolor{red}{#1}}
\renewcommand{\rev}[1]{#1}

\usepackage{amsmath}
\allowdisplaybreaks[4] 

\usepackage[T1]{fontenc} 
\usepackage{tensor}
\usepackage{braket}  
\usepackage{bm}      
\usepackage{upgreek}
\usepackage{extarrows}

\begin{document}
\title{Localized Steps toward ACT-Favored Inflation}

\author{Kai-Ge Zhang}
\email{zhangkaige21@mails.ucas.ac.cn}
\affiliation{Department of Physics, College of Sciences, Shihezi University,
Shihezi 832000, China}

\author{Chengjie Fu}
\email{fucj@ahnu.edu.cn}
\affiliation{Department of Physics, Anhui Normal University, Wuhu, Anhui 241002, China}

\author{Jian-Feng He}
\email{hejianfeng@itp.ac.cn}
\affiliation{Institute of
Theoretical Physics, Chinese Academy of Sciences (CAS), Beijing 100190, China}
\affiliation{School of Physical Sciences, University of Chinese Academy of
Sciences, No.19A Yuquan Road, Beijing 100049, China}

\author{Zong-Kuan Guo}
\email{guozk@itp.ac.cn}
\affiliation{Institute of
Theoretical Physics, Chinese Academy of Sciences (CAS), Beijing 100190, China}
\affiliation{School of Physical Sciences, University of Chinese Academy of
Sciences, No.19A Yuquan Road, Beijing 100049, China}
\affiliation{School of Fundamental Physics and Mathematical Sciences,
Hangzhou Institute for Advanced Study, University of Chinese Academy of
Sciences, Hangzhou 310024, China}


\begin{abstract}
Recent ACT measurements favor a scalar spectral index larger than the Planck value, posing a challenge to many single-field slow-roll inflation models. We show that a smooth, localized step in the inflaton potential can shift the predicted scalar spectral index and tensor-to-scalar ratio by displacing the field value at which the CMB pivot scale exits the horizon. This mechanism can move monomial and, in particular, plateau-like attractor models toward the ACT-favored region, whereas the induced shift remains insufficient in natural inflation. We derive semi-analytical expressions for the step-induced remapping and quantify the associated effective e-fold shift, finding that it can be comparable to, and in some cases exceed, the shift allowed by conservative reheating uncertainties.
\end{abstract}

\maketitle

\section{Introduction}
Inflation is the leading paradigm ~\cite{Brout:1977ix,Starobinsky:1980te,Kazanas:1980tx,Sato:1980yn,Guth:1980zm,Linde:1981mu,Albrecht:1982wi,Linde:1983gd} for the very early Universe, providing a unified explanation of the horizon and flatness puzzles while offering a causal mechanism for the origin of cosmic structure. In its simplest realization, quantum fluctuations of a single slowly rolling scalar field generate an almost scale-invariant curvature power spectrum~\cite{Mukhanov:1981xt,Mukhanov:1982nu,Hawking:1982cz,Guth:1982ec,Starobinsky:1982ee,Bardeen:1983qw,Kodama:1984ziu,Mukhanov:1985rz},
$\mathcal{P}_{\mathcal{R}}(k) = A_{s} (k / k_{\ast})^{n_{s} - 1}$
with the tilt tightly constrained by CMB temperature and polarization data. The Planck 2018 analysis~\cite{Planck:2018vyg,Planck:2018jri} results favor a red-tilted spectrum, $n_s=0.9649\pm0.0042\ (68\%\ \mathrm{C.L.})$.
More importantly, successful inflation must persist long enough to solve the horizon problem and to ensure that the CMB pivot scale exits the horizon roughly \(N_{\ast} \equiv \ln(a_{\rm e}/a_{\ast}) \sim 50\text{--}60\) $e$-folds before the end of inflation, assuming standard post-inflationary evolution~\cite{Liddle:2003as}, where \(a_{\ast}\) and \(a_{\rm e}\) are the scale factors at pivot-scale horizon exit and at the end of inflation, respectively. Together, these requirements favor a sustained slow-roll phase, typically realized by flat, concave potentials approaching a plateau over the field range probed by the CMB. Well-motivated examples include higher-curvature (Starobinsky-type~\cite{Starobinsky:1980te,Vilenkin:1985md}) inflation, Higgs inflation with a non-minimal coupling to gravity~\cite{Bezrukov:2007ep}, and supergravity inspired $\alpha$-attractor models \cite{Kallosh:2013hoa,Kallosh:2013daa,Ferrara:2013rsa,Kallosh:2013yoa,Linde:2015uga}.

However, this picture has been sharpened by recent high-resolution CMB measurements \cite{AtacamaCosmologyTelescope:2025blo,AtacamaCosmologyTelescope:2025nti}. In particular, the combination of Planck, ACT, and Dark Energy Spectroscopic Instrument (DESI) data yields a spectral index of $n_s=0.9743 \pm 0.0034$~\cite{AtacamaCosmologyTelescope:2025blo}. Such a shift places stress on many previous well-performing single-field slow-roll models, because a larger $n_s$ generically requires the pivot scale $k_{\ast}$ to exit the horizon deeper in the slow-roll regime, where the potential is effectively flatter and the slow-roll parameters are smaller. For the broad class of plateau-like potentials, this typically corresponds to shifting the field value $\phi_{\ast}$, at the time when the scale $k_{\ast}$ exits the horizon, further into the asymptotic plateau. In the slow-roll approximation, for the inflationary potential $V(\phi)$, $N_{\ast} \simeq \int_{\phi_{\rm e}}^{\phi_{\ast}} \bigl(V/V^{\prime}\bigr) \, d\phi / M_{\mathrm{pl}}^{2}$, where $\phi_{\rm e}$ denotes the field value at the end of inflation. Shifting $\phi_{\ast}$ to a flatter region is therefore generally associated with larger $N_{\ast}$. This exposes a tension between the ACT-preferred tilt and a conventional window $N_{\ast}$ in minimal single-field scenarios.

Several directions have been explored to alleviate this tension. Beyond the minimal Einstein-frame single-field slow-roll framework, modifications such as higher-curvature or modified-gravity terms~\cite{Gialamas:2025ofz,Yogesh:2025wak,Ketov:2025cqg,Addazi:2025qra,Odintsov:2025bmp,Zhu:2025twm,zhang2025act,fu2025harrison}, nonminimal couplings and Palatini realizations~\cite{Dioguardi:2025vci,Gao:2025onc,Dioguardi:2025mpp}, as well as noncanonical kinetic structures~\cite{He:2025bli,Gao:2025viy}, can modify the inflationary dynamics and shift plateau-model predictions toward larger values of $n_s$.
Alternatively, non-instantaneous reheating~\cite{zharov2025reheating,haque2025act,liu2025reconciling,gialamas2026reheating}, including mechanisms such as infrared gravitational reheating~\cite{Chakraborty:2025oyj} or preheating~\cite{risdianto2025preheating}, can modify the post-inflationary expansion history and hence shift the number of $e$-folds associated with horizon exit, thereby improving agreement with ACT. Such scenarios, however, typically rely on specific assumptions about the reheating microphysics and are subject to BBN and $\Delta N_{\rm eff}$ constraints~\cite{liu2025reconciling,Chakraborty:2025oyj,Drees:2025ngb,haque2025act,Maity:2025czp}.
Explicit radiative corrections have also been reconsidered: one-loop effects can reshape the effective potential and thereby shift $n_s$ into the ACT-preferred range~\cite{Gialamas:2025kef,Wolf:2025ecy}, but typically in a way that is sensitive to the UV completion and the choice of renormalization scheme.



By contrast, in this work we pursue a minimal route to easing this tension: as a phenomenological parametrization of a localized feature applicable across different potential forms, we consider a step-like modulation of the inflaton potential~\cite{Starobinsky:1992ts,Adams:2001vc,Bartolo:2013exa}, while retaining the canonical single-field framework and leaving both the gravitational sector and the post-inflationary history unchanged. We first analyze how the step reshapes the background evolution, most notably the mapping between the pivot field value $\phi_{\ast}$ and the $e$-folding number $N_{\ast}$. We then confront several representative deformed single-field models with current CMB constraints on $(n_s, r)$, and show that the step can accommodate the ACT-preferred tilt while maintaining a viable $e$-fold 
range. \rev{We further derive semi-analytical expressions for the step-induced remapping and use them to interpret the numerical results. Finally, we compare the effective e-fold shift generated by the step with the uncertainty associated with reheating, showing that a localized feature can produce a reheating-scale shift without modifying the post-inflationary expansion history.}

\section{Step-Modulated Single-Field Inflation}
\label{sec: MODEL}
We work in the Einstein frame and consider a canonical scalar field minimally coupled to gravity,
\begin{equation}
S=\int d^{4}x\,\sqrt{-g}\left[\frac{M_{\mathrm{pl}}^{2}}{2}R-\frac{1}{2}g^{\mu\nu}\partial_{\mu}\phi\,\partial_{\nu}\phi - V(\phi)\right],
\label{eq:action}
\end{equation}
where $M_{\mathrm{pl}}$ is the reduced Planck mass and $\phi$ denotes the inflaton. 
In a spatially flat Friedmann-Robertson-Walker background with line element ${\rm d}s^2=-{\rm d}t^2 + a(t)^2\delta_{ij}{\rm d}x^i{\rm d}x^j$, 
the inflationary dynamics are governed by
\begin{align}
\label{BG1}
&3M_{\mathrm{pl}}^2 H^2 = \frac{1}{2}\dot\phi^2 + V_{0}(\phi)\xi(\phi), \\
&\ddot\phi + 3H\dot\phi + V_{0}^{\prime}(\phi)\xi(\phi)
+ V_{0}(\phi)\xi^{\prime}(\phi) = 0,
\end{align}
where $H$ is the Hubble parameter, and the potential is written as
$V(\phi)=V_0(\phi)\,\xi(\phi)$. Here \(V_0(\phi)\) denotes the underlying single-field slow-roll potential, such as  the \(\alpha\)-attractor models, and \(\xi(\phi)\) is a smooth step modulation,
\begin{equation}
\xi(\phi)=1+\gamma\,\tanh\!\left(\frac{\phi-\phi_c}{\Delta\phi}\right),
\label{eq:xi_step}
\end{equation}
where \(\gamma\) and \(\Delta\phi\) control the step height and width, respectively. The strength and sharpness of the feature are characterized by
\begin{align}
\frac{\xi'}{\xi}&=\frac{\gamma}{\Delta\phi}\,
\frac{\operatorname{sech}^{2}x}{1+\gamma\tanh x},
\qquad
\frac{\xi''}{\xi}=-\frac{2\gamma}{\Delta\phi^{2}}\,
\frac{\operatorname{sech}^{2}x\,\tanh x}{1+\gamma\tanh x},
\label{eq:xi_derivs}
\end{align}
where \(x\equiv (\phi-\phi_c)/\Delta\phi\). Throughout, we take \(|\gamma|\lesssim \mathcal{O}(0.1)\) and \(\Delta\phi \ll |\phi_\ast-\phi_{\rm e}|\), so that the deformation remains localized and perturbative. To ensure that the step modifies only the background evolution, rather than generating a sharp feature~\cite{Dvorkin:2009ne,Adams:2001vc,Bartolo:2013exa}, we further require
\begin{equation}
M_{\mathrm{pl}}\left|\frac{\xi'}{\xi}\right|
\lesssim 1,
\qquad
M_{\mathrm{pl}}^{2}\left|\frac{\xi''}{\xi}\right|\lesssim 1.
\label{eq:step_sr_cond}
\end{equation}
For a given step height \(\gamma\), these conditions imply a lower bound on the width \(\Delta\phi\).

Under Eq.~\eqref{eq:step_sr_cond}, the inflationary dynamics admit a simple analytic description in the slow-roll approximation,
\begin{align}
\label{eq:bg_nr}
&3M_{\mathrm{pl}}^2 H^2 \simeq V_{0}(\phi)\xi(\phi),\\
&3H\dot\phi + V_{0}^{\prime}(\phi)\xi(\phi)
+ V_{0}(\phi)\xi^{\prime}(\phi) \simeq 0.
\label{eq:bg_sr}
\end{align}
These equations clearly show how the localized step alters the inflaton dynamics relative to those in the undeformed, nearly flat background potential \(V_0(\phi)\).

During the first stage of the evolution, well before the inflaton reaches the step, we identify the CMB field interval \(\mathcal{I}_{\rm CMB}\) with that of the undeformed model and require it to lie well outside the step region,
\begin{equation}
\min_{\phi\in \mathcal{I}_{\rm CMB}}
\left|\frac{\phi-\phi_c}{\Delta\phi}\right| \gg 1 .
\label{eq:CMB_far_step}
\end{equation}
For fixed \(\Delta\phi\) and \(\mathcal{I}_{\rm CMB}\), this condition primarily constrains the step location \(\phi_c\). In this regime, \(\xi(\phi)\) is effectively constant over \(\mathcal{I}_{\rm CMB}\), and its derivatives are exponentially suppressed. The CMB-scale dynamics are therefore indistinguishable from those of the undeformed model, implying
\begin{align}
\epsilon_V
&\equiv \frac{M_{\mathrm{pl}}^2}{2}\left(\frac{V'}{V}\right)^2
\simeq
\frac{M_{\mathrm{pl}}^2}{2}\left(\frac{V_0'}{V_0}\right)^2,\\
\eta_V
&\equiv M_{\mathrm{pl}}^2\frac{V''}{V}
\simeq
M_{\mathrm{pl}}^2\frac{V_0''}{V_0},
\end{align}
so that
\begin{equation}
n_s-1 \simeq -6\epsilon_V+2\eta_V,\qquad r\simeq 16\epsilon_V,
\label{eq:nsr}
\end{equation}
remain unchanged at leading order.

Once the inflaton enters the transition region, \(|\phi-\phi_c|\lesssim \Delta\phi\), the evolution temporarily deviates from that of the undeformed model. For definiteness, we first consider plateau-like potentials in which the inflaton rolls from larger to smaller field values, so that $V_{0}'/V_{0} > 0$. The case with the opposite rolling direction is entirely analogous, with the relevant sign conventions reversed. In this case, the leading effect of the step is to modify the local logarithmic derivative of the potential,
\begin{equation}
\frac{V^{\prime}}{V} = \frac{V_{0}^{\prime}}{V_{0}} + \frac{\xi^{\prime}}{\xi},
\label{eq:VVprime_split}
\end{equation}
thereby changing the number of \(e\)-folds accumulated across the transition. The resulting local shift is
\begin{equation}
\delta N \equiv \int_{\phi_{c}+l\Delta\phi}^{\phi_{c}-l\Delta\phi}
\left(\frac{V}{M_{\mathrm{pl}}^2V'}-\frac{V_0}{M_{\mathrm{pl}}^2V_0'}\right)\,{\rm d}\phi,
\label{eq:delta_N}
\end{equation}
where \(l=\mathcal{O}(1)\) is chosen so that \(V'/V\sim V_0'/V_0\) outside the integration range. Across the transition, Eq.~\eqref{eq:delta_N} can be rewritten as
\begin{equation}
\delta N \simeq
-\frac{1}{M_{\mathrm{pl}}^2}
\int_{\phi_c+l\Delta\phi}^{\phi_c-l\Delta\phi}
\left(\frac{V_0}{V_0'}\right)^2\frac{\xi'}{\xi}
\left(1+\frac{V_0}{V_0'}\frac{\xi'}{\xi}\right)^{-1}
\,{\rm d}\phi .
\label{eq:deltaN_master}
\end{equation}
This expression provides a convenient starting point for analyzing the steepening and flattening limits discussed below. 

If the step steepens the potential, \(\gamma>0\), so that \(V^{\prime}/V>V_0^{\prime}/V_0\), the inflaton crosses the same field interval more rapidly than in the undeformed model, and fewer \(e\)-folds are accumulated across the step region. A sizable reduction requires
\begin{equation}
\frac{\xi'}{\xi}\gg \frac{V_0'}{V_0}.
\label{eq:step_dominates}
\end{equation}
Expanding Eq.~\eqref{eq:deltaN_master} in this step-dominated regime gives
\begin{equation}
\delta N \simeq
\mathcal{O}(1)\frac{\Delta\phi}{M_{\mathrm{pl}}^{2}}
\left(\frac{V_0}{V_0'}\right)_{\phi_c},
\label{eq:deltaN_steepen_result}
\end{equation}
which is manifestly positive.

By contrast, if the step locally flattens the potential, \(\gamma<0\), so that \(V^{\prime}/V<V_0^{\prime}/V_0\), the inflaton crosses the same field interval more slowly than in the undeformed model, thereby accumulating more \(e\)-folds across the step region. To avoid a reversal of the field velocity, and hence the unphysical trapping of the inflaton, one must require
\begin{equation}
-\frac{\xi'}{\xi}< \frac{V_0'}{V_0}.
\label{eq:no_trap}
\end{equation}
In this regime, expanding Eq.~\eqref{eq:deltaN_master} to leading order yields
\begin{equation}
\delta N \simeq
\frac{1}{M_{\mathrm{pl}}^{2}}
\left(\frac{V_0}{V_0'}\right)_{\phi_c}^{2}
\ln\!\frac{1+\gamma}{1-\gamma},
\label{eq:deltaN_flatten_result}
\end{equation}
which is negative.

Finally, after the inflaton has rolled a few widths beyond the transition, \(\xi(\phi)\) again approaches a constant, and the subsequent evolution becomes approximately identical to that of the undeformed model. The localized step therefore leaves a net shift \(\delta N\) in the \(e\)-folds accumulated across the transition, as quantified by Eq.~\eqref{eq:delta_N}.

\begin{figure}[htb]
\centering
\includegraphics[width=1\columnwidth]{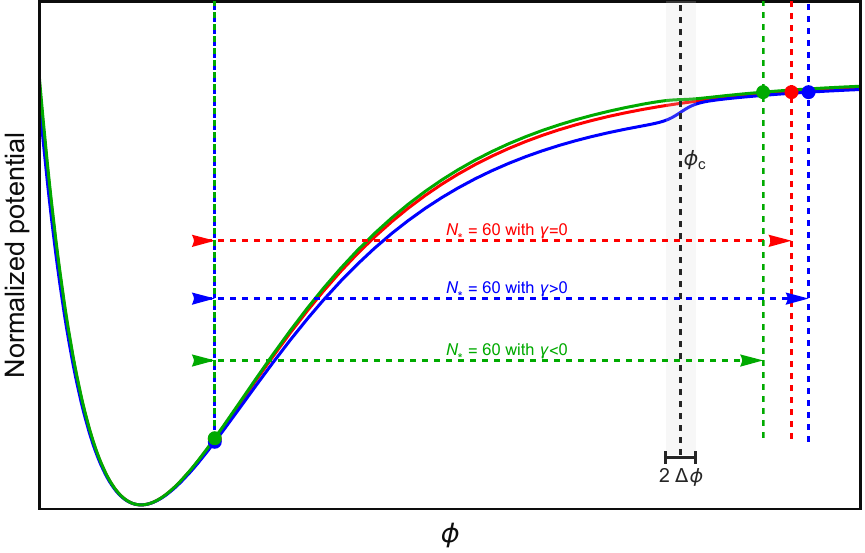}
\caption{\label{fig1}
Starobinsky potentials with a localized step feature for $\gamma=0$, $\gamma>0$, and $\gamma<0$, normalized at their respective $\phi_\ast$. Colored dashed lines and dots indicate the corresponding $\phi_\ast$ and $\phi_{\rm e}$ at fixed $N_\ast=60$.
}
\end{figure}

For fixed total duration of inflation, e.g. \(N_\ast=60\), this shift must be compensated by the pre-transition stage containing the CMB pivot scale. The field value at which the pivot scale exits the horizon is then shifted from \(\phi_\ast^{(0)}\) to \(\phi_\ast\), such that
\begin{equation}
\Delta N_{\rm comp} \simeq -\,\delta N,
\label{eq:compensation_condition}
\end{equation}
where
\begin{equation}
\Delta N_{\rm comp} \equiv
\int_{\phi_\ast}^{\phi_\ast^{(0)}}
\frac{V_0}{M_{\mathrm{pl}}^{2}V_0'}\,{\rm d}\phi,
\label{eq:Delta_N}
\end{equation}
and \(\phi_\ast^{(0)}\) denotes the corresponding field value in the undeformed model.
The step therefore redistributes the fixed {\it e}-fold budget and, in doing so, shifts the field value at which the pivot scale exits the horizon, as illustrated in Fig.~\ref{fig1}. For \(\gamma>0\), one typically has \(\delta N>0\), and hence \(\Delta N_{\rm comp}<0\), so $\phi_\ast$ shifts to a larger value, toward a flatter region of the potential. For \(\gamma<0\), one typically has \(\delta N<0\), hence \(\Delta N_{\rm comp}>0\), so $\phi_\ast$ shifts to a smaller value, where the potential is steeper. The prediction is therefore displaced along the undeformed attractor trajectory in the \((n_s,r)\) plane: positive steps generally move it toward larger \(n_s\) and smaller \(r\), while negative steps shift it in the opposite direction. 

\begin{figure}[htb]
\centering
\includegraphics[width=1\columnwidth]{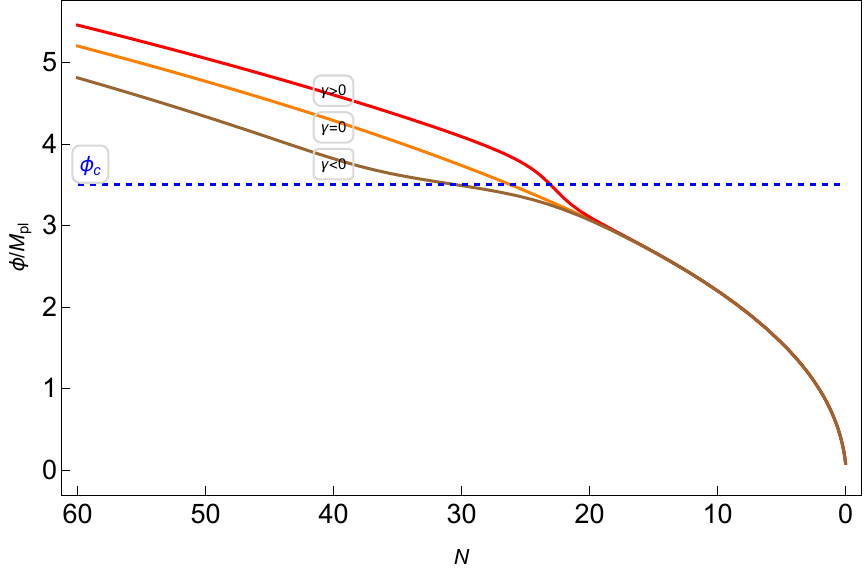}
\caption{\label{fig2} Evolution of $\phi$ with the number of $e$-folds $N$ for the monomial potential with $p=1/10$, shown for $\gamma<0$, $\gamma=0$, and $\gamma>0$.}
\end{figure}

\begin{figure*}[t]
    \centering
    \includegraphics[width=0.8\textwidth]{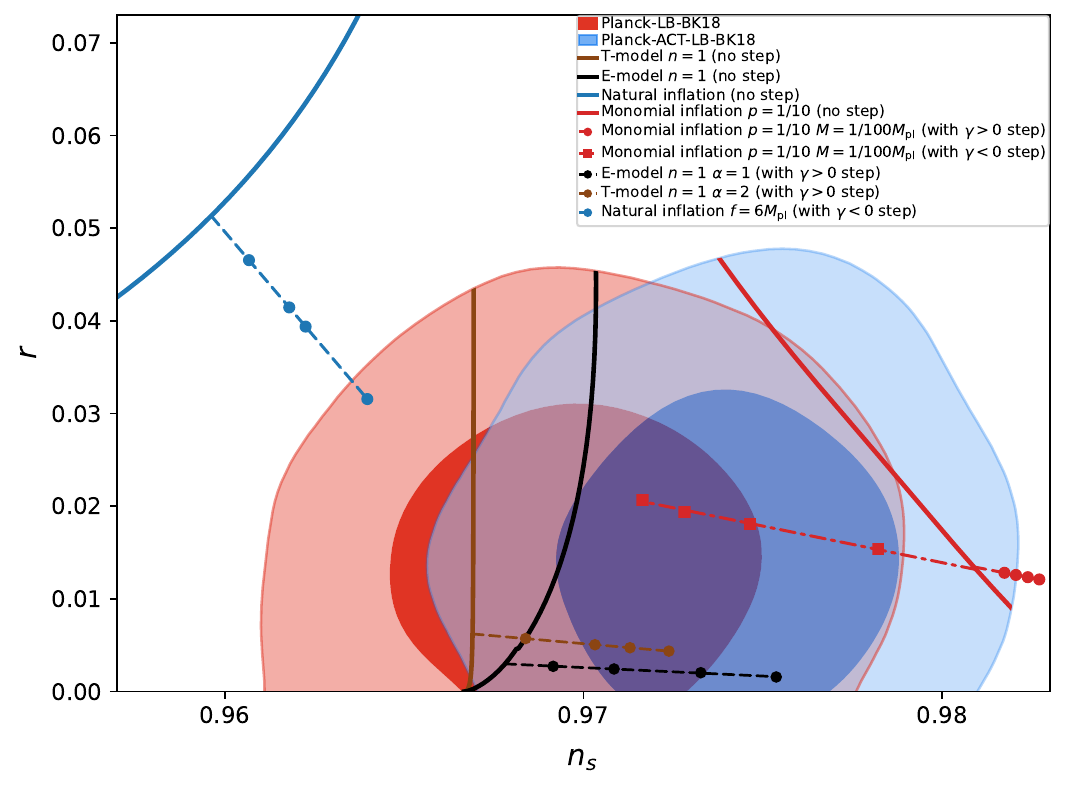}
\caption{
Theoretical predictions in the $(n_s,r)$ plane for representative single-field
inflation models, with and without a localized step in the potential.
The red shaded region shows the constraint from the Planck-LB-BK18 combination~\cite{AtacamaCosmologyTelescope:2025nti}, where LB denotes the inclusion of CMB lensing~\cite{ACT:2023kun,Carron:2022eyg} and BAO data~\cite{DESI:2024uvr,DESI:2024mwx}, and BK18 refers to the BICEP/Keck 2018 likelihood~\cite{BICEP:2021xfz}. The blue shaded region shows the corresponding constraint from the Planck-ACT-LB-BK18 combination~\cite{AtacamaCosmologyTelescope:2025nti}, which further includes ACT data~\cite{AtacamaCosmologyTelescope:2025blo}, with lensing and BAO information taken from ACT+Planck~\cite{ACT:2023kun,Carron:2022eyg} and DESI~\cite{DESI:2024uvr,DESI:2024mwx}, respectively. Solid curves without markers show the predictions of the unmodified models at $N_{*}=60$: monomial inflation with $p=1/10$ (red), the $\alpha$-attractor T-model with $n=1$ (brown), the $\alpha$-attractor E-model with $n=1$ (black), and natural inflation (blue). Colored curves and filled markers show the corresponding predictions with a step modulation. For the monomial potential, $p=1/10$ and $M=1/100M_{\mathrm{pl}}$ are fixed; the two parameter sets are $(\gamma,\Delta\phi/M_{\mathrm{pl}})=(0.025,0.143)$ with $\phi_{c}/M_{\mathrm{pl}}=\{4.5,4.0,3.5,3.0\}$ (red circles), and $(\gamma,\Delta\phi/M_{\mathrm{pl}})=(-0.008,0.143)$ with $\phi_{c}/M_{\mathrm{pl}}=\{3.66,3.6,3.5,3.0\}$ (red squares). For the T-model and E-model, $n=1$, $\gamma=0.1$, and $\Delta\phi=0.286M_{\mathrm{pl}}$ are fixed; the representative choices are $\alpha=2$ with $\phi_{c}/M_{\mathrm{pl}}=\{6.2,5.9,5.6,4.8\}$ (brown circles) and $\alpha=1$ with $\phi_{c}/M_{\mathrm{pl}}=\{4.5,4.0,3.5,3.0\}$ (black circles). For natural inflation, $f=6\,M_{\mathrm{pl}}$, $\gamma=-0.5$, and $\Delta\phi=1.0M_{\mathrm{pl}}$, with $\phi_{c}/M_{\mathrm{pl}}=\{14.0,12.0,11.3,9.0\}$ (blue circles). All benchmark choices satisfy the separation condition in Eq.~\eqref{eq:CMB_far_step} and the controlled-deviation conditions in Eq.~\eqref{eq:step_sr_cond}.
}
\label{fig3}
\end{figure*}


\section{Application to ACT-favored models}

\label{sec: Background Dynamics}
In this section, we apply the step-modulated setup to several representative classes of single-field inflation models and assess whether the resulting predictions can be brought into agreement with current observations. In all cases, we impose the CMB constraints on $n_s$ and $r$ in Eq.~\eqref{eq:nsr} from Planck, BICEP/Keck, and ACT, adopting a pivot scale of $k_\ast=0.05\,\mathrm{Mpc}^{-1}$.

We first consider monomial inflation~\cite{Damour:1997cb}, described by
\begin{align}\label{monomial_potentials}
    V_0(\phi)=\Lambda_{\rm mon}^4\left[\left(1+\frac{\phi^2}{M^2}\right)^{p/2}-1\right],
\end{align}
which is strongly disfavored by the combined Planck and BICEP/Keck constraints, but becomes less discrepant with the higher scalar tilt preferred by recent ACT data. As an illustrative example, we use this potential to examine how the sign of the step modulation, $\gamma < 0$ or $\gamma > 0$, affects the inflationary dynamics. Throughout this example, we fix $p = 1/10$, $M = 1/100 M_{\mathrm{pl}}$, $\phi_{c} = 3.5 M_{\mathrm{pl}}$, and $\Delta\phi = 0.25 M_{\mathrm{pl}}$, and consider $\gamma = -0.01$ and $\gamma = 0.025$. The resulting evolution of $\phi$ as a function of $N$ is shown in Fig.~\ref{fig2}. Relative to the undeformed case $\gamma = 0$, the field interval from $\phi_{c}$ to $\phi_{e}$ corresponds to fewer {\it e}-folds for $\gamma > 0$ and to more {\it e}-folds for $\gamma < 0$. At fixed $N_{*} = 60$, this shifts the field value $\phi_{*}$ to larger values for $\gamma > 0$ and to smaller values for $\gamma < 0$. As illustrated schematically in Fig.~\ref{fig3}, the step can shift $\phi_{*}$ toward either a flatter or a steeper region of the potential. Which direction is phenomenologically favored depends on the class of models under consideration. For monomial inflation, whose scalar tilt at $N_{*} = 60$ is already relatively large, the preferred shift is toward a steeper region. By contrast, for the plateau-like models considered below, whose undeformed predictions typically have smaller $n_{s}$, the preferred shift is toward a flatter region. Since the inflaton may roll toward either smaller or larger field values, the corresponding sign of $\gamma$ is model dependent.

We next consider a class of plateau inflation models described by the potentials
\begin{equation}
V_0(\phi)=\Lambda_{\rm E}^4\left(1-e^{-\sqrt{\frac{2}{3\alpha}}\,\frac{\phi}{M_{\mathrm{pl}}}}\right)^{2},
\end{equation}
and
\begin{equation}
V_0(\phi)=\Lambda_{\rm T}^4\tanh^{2}\!\left(\frac{\phi}{\sqrt{6\alpha}M_{\mathrm{pl}}}\right),
\end{equation}
which belong to the class of $\alpha$-attractor models~\cite{Kallosh:2013yoa,Kallosh:2014rga}.
These potentials feature an asymptotically flat plateau at large field values and lead to universal predictions for the scalar spectral index,
\begin{equation}
n_s \simeq  1 - \frac{2}{N},
\end{equation}
which is largely insensitive to the detailed form of the potential.
As a result, they are in good agreement with the Planck and BICEP/Keck observations~\cite{BICEP:2021xfz} for a wide range of model parameters. The inclusion of recent ACT data, however, favors a slightly larger scalar spectral index, thereby significantly narrowing the viable parameter space compared to that inferred from Planck alone. In particular, for a fixed number of $e$-folds, $N_\ast=60$, the predictions of
the undeformed potentials lie near the lower edge of the ACT-allowed region in the $(n_s,r)$ plane.
Within the step-modulated framework considered here, the total number of $e$-folds is held fixed, while a positive step ($\gamma>0$) shifts the prediction toward larger $n_s$ and smaller $r$, as shown in Fig.~\ref{fig3}. This shift improves agreement with the combined Planck--ACT--LB--BK18 constraints, consistent with the mechanism described by Eq.~\eqref{eq:delta_N} and Eqs.~\eqref{eq:compensation_condition}--\eqref{eq:Delta_N}.

Finally, we consider natural inflation~\cite{Freese:1990rb},
\begin{equation}
V_0(\phi)=\Lambda_{\rm NI}^4\left[1+\cos\!\left(\frac{\phi}{f}\right)\right].
\end{equation}
Unlike the attractor models discussed above, natural inflation remains tied to the hilltop relation,
\begin{equation}
n_s \simeq 1-\frac{M_{\mathrm{pl}}^2}{f^2},
\end{equation}
so the observed tilt already requires \(f\gtrsim \mathcal{O}(5)\,M_{\mathrm{pl}}\). 
At fixed $N_{*} = 60$, however, natural inflation retains its familiar trade-off: smaller $f$ makes the spectrum too red, whereas larger $f$ drives the model toward the quadratic limit and hence larger $r$. The model is therefore strongly disfavored by Planck and remains well outside the ACT-preferred region in the \((n_s,r)\) plane. A negative step, \(\gamma<0\), can shift the prediction in the right direction, but only modestly. As shown in Fig.~\ref{fig3}, even for \(f=6M_{\mathrm{pl}}\) and a relatively broad step, \(\Delta\phi=1.0M_{\mathrm{pl}}\), the prediction still lies outside the ACT-favored region. Increasing \(\Delta\phi\) further would spoil localization on the characteristic field scale \(\sim \pi f\), while moving \(\phi_c\) closer to the observable window would violate the separation condition in Eq.~\eqref{eq:CMB_far_step} and thus affect CMB scales. We have verified numerically that varying \(f\) does not qualitatively change this conclusion.


\section{Semi-analytical estimate of the step-induced remapping}
\label{sec: Semi-analytical calculation}
In this section, we derive semi-analytical expressions for the step-induced remapping of $\phi_\ast$, providing an analytic understanding of the shifts in $n_s$ and $r$ found in Sec.~\ref{sec: Background Dynamics}. The $e$-folding number from the time $t_*$ at which $\phi = \phi_*$ to the time $t_{\rm e}$ at the end of inflation is defined by
\begin{equation}
N_* = \int_{t_*}^{t_{\rm e}} H(t) \, \mathrm{d}t.
\end{equation}
Using Eqs.~\eqref{eq:bg_nr} and ~\eqref{eq:bg_sr} in the main text, we obtain
\begin{align}
N_* &= \frac{1}{M_{\mathrm{pl}}^2} \int_{\phi_{\rm e}}^{\phi_*}  \frac{V}{V^{\prime}} \, \mathrm{d}\phi 
\nonumber \\
&\simeq \frac{1}{M_{\mathrm{pl}}^2} \!\!\int_{\phi_{\rm e}}^{\phi_*} \frac{V_0}{V_0^{\prime}}  \mathrm{d}\phi +  \frac{1}{M_{\mathrm{pl}}^2} \!\!\int_{\phi_{\rm c}-\Delta\phi}^{\phi_{\rm c}+\Delta\phi} \left[\left(\frac{V_{0}^{\prime}}{V_{0}} + \frac{\xi^{\prime}}{\xi}\right)^{-1} \!\!- \frac{V_0}{V_0'}\right]\mathrm{d}\phi\nonumber \\
\label{eq:A1}
\end{align}
where $\phi_{\rm e}$ denotes the field value at $t = t_{\rm e}$. 

For the semi-analytical estimates below, we neglect the small end-point correction and take $\phi_{\rm e}$ to be the minimum of the corresponding undeformed potential, e.g., $\phi_{\rm e} = 0$ for the monomial potential and the $\alpha$-attractor E-model. Using the local-slope approximation in the transition region, 
\begin{equation}
\frac{V_0'(\phi)}{V_0(\phi)} \simeq \left. \frac{V_0'}{V_0} \right|_{\phi_c},
\label{eq:V0_slope_approx}
\end{equation}
Eq.~\eqref{eq:A1} can be integrated analytically. The relevant expressions for the representative models are given below and are written compactly in terms of the dimensionless local slope
\begin{equation}
a_c \equiv \left. \frac{V_0'}{V_0} \right|_{\phi_c}\Delta\phi .
\end{equation}

\begin{figure}[htb]
\centering
\includegraphics[width=1\columnwidth]{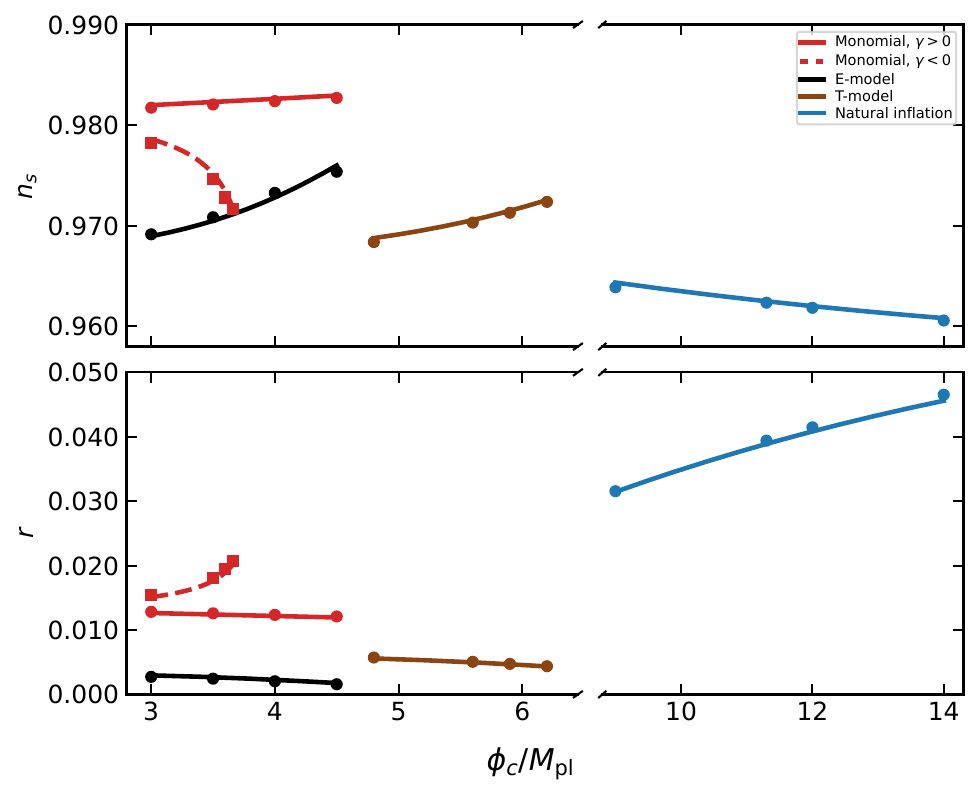}
\caption{\rev{The (semi-)analytical predictions for $n_s$ and $r$ as functions of $\phi_c$ for monomial potential with $\{p = 1/10, M = 1/100M_{\mathrm{pl}}, \gamma=0.025, \Delta\phi/M_{\mathrm{pl}}=0.143\}$ (red solid line) and $\{p = 1/10, M = 1/100M_{\mathrm{pl}},\gamma=-0.008, \Delta\phi/M_{\mathrm{pl}}=0.143\}$ (red dashed line), $\alpha$-attractor E-model with $\{\alpha = 1, \gamma=0.1, \Delta\phi/M_{\mathrm{pl}}=0.286\}$ (black solid line), $\alpha$-attractor T-model with $\{\alpha = 2, \gamma=0.1, \Delta\phi/M_{\mathrm{pl}}=0.286\}$ (brown solid line), and natural inflation with $\{f = 6M_{\mathrm{pl}}, \gamma=-0.5, \Delta\phi/M_{\mathrm{pl}}=1.0\}$ (blue solid line). The dots and squares correspond to the numerical results shown in Fig.~\ref{fig3}.}}
\label{fig4}
\end{figure}

Fig.~\ref{fig4} shows the analytical predictions for $n_s$ and $r$ as functions of $\phi_{c}$ for these four potentials. One can readily see that the analytical results are in good agreement with the numerical ones.

\subsection{Monomial inflation}

We first derive the semi-analytical predictions for the step-modulated monomial model, for which the e-folding number becomes
\begin{align}
N_* &\simeq   \frac{\phi_*^2}{2pM_{\mathrm{pl}}^2} \left[ 1 + \frac{2}{p-2} \left( \frac{\phi_*}{M} \right)^{-p} \right] 
 -\frac{\Delta \phi^2}{ a_c \sqrt{1+a_c/\gamma} M_{\mathrm{pl}}^2} \nonumber \\ &\times \left[\operatorname{arctanh}\left( \frac{2-a_c}{2\sqrt{1+a_c/\gamma}} \right)-\operatorname{arctanh}\left( \frac{-2-a_c}{2\sqrt{1+a_c/\gamma}} \right)\right].
 \label{eq:A2}
\end{align} 

The scalar spectral index and the tensor-to-scalar ratio at the pivot scale are given by
\begin{align}
n_s&\simeq1-
\left[
p(p+2)\left(\frac{\phi_*}{M}\right)^{2p}
+2p(p-1)\left(\frac{\phi_*}{M}\right)^p
\right]\nonumber \\&\times
\left[
\left(\frac{\phi_*}{M}\right)^p-1
\right]^{-2}
\left(\frac{\phi_*}{M_{\mathrm{pl}}}\right)^{-2},
\label{eq:A3}
\end{align}
and
\begin{equation}
r\simeq
8p^2
\left(\frac{\phi_*}{M}\right)^{2p}
\left[
\left(\frac{\phi_*}{M}\right)^p-1
\right]^{-2}
\left(\frac{\phi_*}{M_{\mathrm{pl}}}\right)^{-2}.
\label{eq:A4}
\end{equation}
Semi-analytical predictions for $n_s$ and $r$ can then be obtained by numerically solving Eq.~\eqref{eq:A2} and substituting the resulting $\phi_*$ into Eqs.~\eqref{eq:A3} and~\eqref{eq:A4}.

\subsection{$\alpha$-attractors}
For the $\alpha$-attractor E-model, one finds
\begin{align}
N_* &\simeq \frac{3\alpha}{4} \left[ \exp\left( \sqrt{\frac{2}{3\alpha}} \frac{\phi_*}{M_{\mathrm{pl}}} \right) - 1 \right]\nonumber \\ &-\frac{\Delta \phi^2}{M_{\mathrm{pl}}^2 a_c \sqrt{1+a_c/\gamma}} \operatorname{arctanh}\left( \frac{2\sqrt{1+a_c/\gamma}}{2+a_c/\gamma} \right)
\end{align}
which yields
\begin{align}
\frac{\phi_*}{M_{\mathrm{pl}}} &\simeq \sqrt{\frac{3\alpha}{2}} \ln \left[ 1 + \frac{4N_*}{3\alpha}  + \frac{4\Delta \phi^2}{3\alpha  a_c \sqrt{1+a_c/\gamma} M_{\mathrm{pl}}^2} \right.\nonumber \\ &\times \left.\operatorname{arctanh}\left( \frac{2\sqrt{1+a_c/\gamma}}{2+a_c/\gamma} \right)\right].
\end{align}

Substituting the resulting $\phi_*$ into the expressions for $n_s$ and $r$, we obtain
\begin{align}
n_s & \simeq 1 - \frac{8}{3\alpha}e^{-\sqrt{\frac{2}{3\alpha}}\frac{\phi_*}{M_{\mathrm{pl}}}}  \nonumber \\& \simeq  1 - \frac{8}{
    4N_* + 3\alpha + \frac{4\Delta \phi^2}{ a_c \sqrt{1+a_c/\gamma} M_{\mathrm{pl}}^2} \operatorname{arctanh}\left( \frac{2\sqrt{1+a_c/\gamma}}{2+a_c/\gamma} \right).
}, \\
r &= \frac{64}{3\alpha \left(e^{\sqrt{\frac{2}{3\alpha}}\frac{\phi_*}{M_{\mathrm{pl}}}} - 1\right)^2}  \nonumber \\& \simeq \frac{192\alpha}{\left[
    4N_* + \frac{4\Delta \phi^2}{ a_c \sqrt{1+a_c/\gamma} M_{\mathrm{pl}}^2} \operatorname{arctanh}\left( \frac{2\sqrt{1+a_c/\gamma}}{2+a_c/\gamma} \right)
\right]^2}.
\end{align}

We next turn to the $\alpha$-attractor T-model, for which
\begin{align}
N_* &\simeq \frac{3\alpha}{4} \left[ \cosh\left(\sqrt{\frac{2}{3\alpha}}\frac{\phi_*}{M_{\mathrm{pl}}}\right) - 1\right] -\frac{\Delta \phi^2}{M_{\mathrm{pl}}^2 a_c \sqrt{1+a_c/\gamma}} \nonumber \\ &\times\operatorname{arctanh}\left( \frac{2\sqrt{1+a_c/\gamma}}{2+a_c/\gamma} \right).
\end{align}
Solving this equation for $\phi_*$, we obtain
\begin{align}
\frac{\phi_*}{M_{\mathrm{pl}}} &\simeq \sqrt{\frac{3\alpha}{2}} \, \operatorname{arccosh} \left[ 1 + \frac{4N_*}{3\alpha} + \frac{4\Delta \phi^2}{3\alpha  a_c \sqrt{1+a_c/\gamma} M_{\mathrm{pl}}^2} \right. \nonumber \\
&\left.\times  \operatorname{arctanh}\left( \frac{2\sqrt{1+a_c/\gamma}}{2+a_c/\gamma} \right) \right]
\end{align}

The corresponding expressions for $n_s$ and $r$ are
\begin{align}
n_s &= 1 - \frac{8}{3\alpha} \frac{1}{\cosh\left(\sqrt{\frac{2}{3\alpha}}\frac{\phi_*}{M_{\mathrm{pl}}}\right) - 1} \nonumber\\&\simeq 1 - \frac{8}{4N_* + \frac{4\Delta \phi^2}{ a_c \sqrt{1+a_c/\gamma} M_{\mathrm{pl}}^2} \operatorname{arctanh}\left( \frac{2\sqrt{1+a_c/\gamma}}{2+a_c/\gamma} \right)}\\
r &= \frac{64}{3\alpha} \operatorname{csch}^2\left(\sqrt{\frac{2}{3\alpha}}\frac{\phi_*}{M_{\mathrm{pl}}}\right)\nonumber\\&\simeq \frac{192\alpha}{\left[4N_* + 3\alpha + \frac{4\Delta \phi^2}{ a_c \sqrt{1+a_c/\gamma} M_{\mathrm{pl}}^2} \operatorname{arctanh}\left( \frac{2\sqrt{1+a_c/\gamma}}{2+a_c/\gamma} \right)\right]^2}.
\end{align}

\subsection{natural inflation}
Finally, we consider natural inflation, for which
\begin{align}
N_* &\simeq -\frac{2f^2}{M_{\mathrm{pl}}^2} \ln \left[ \sin \left( \frac{\phi_*}{2f} \right) \right] -\frac{\Delta \phi^{2}}{M_{\mathrm{pl}}^{2} a_{c} \sqrt{1-a_c/\gamma}} \nonumber\\&\times\operatorname{arctanh}\left( \frac{2\sqrt{1-a_c/\gamma}}{2-a_c/\gamma} \right)
\end{align}
Solving this equation for $\phi_*$, we find
\begin{align}
\phi_{*} &= 2f \arcsin \left[ \exp \left( -\frac{M_{\mathrm{pl}}^{2}}{2f^{2}} \left( N_{*} + \frac{\Delta \phi^{2}}{M_{\mathrm{pl}}^{2} a_{c} \sqrt{1-a_c/\gamma}} \right.\right.\right.\nonumber\\&\times\left.\left.\left.\operatorname{arctanh}\left( \frac{2\sqrt{1-a_c/\gamma}}{2-a_c/\gamma} \right)\right) \right) \right]
\end{align}

The corresponding expressions for $n_s$ and $r$ are given by 
\begin{equation}
n_s = \frac{1}{2} \left[ 1 - \frac{3M_{\mathrm{pl}}^2}{f^2} + \left( 1 + \frac{M_{\mathrm{pl}}^2}{f^2} \right) \cos \frac{\phi_*}{f} \right] \sec^2 \frac{\phi_*}{2f},
\end{equation}
\begin{equation}
r = \frac{8M_{\mathrm{pl}}^2}{f^2} \tan^2 \frac{\phi_*}{2f}.
\end{equation}

\section{STEP-INDUCED SHIFTS VERSUS REHEATING UNCERTAINTIES}
The localized step considered above changes the inflationary prediction by modifying the mapping between the CMB pivot field value and the e-folding number. A nonstandard post-inflationary history can also shift the prediction in the \((n_s,r)\) plane, since the e-folding number assigned to the CMB pivot scale depends on the expansion between the end of inflation and radiation domination. This motivates a direct comparison between the step-induced e-fold shift and the range allowed by reheating uncertainty.

\begin{figure*}[t]
\centering
\includegraphics[width=0.6\textwidth]{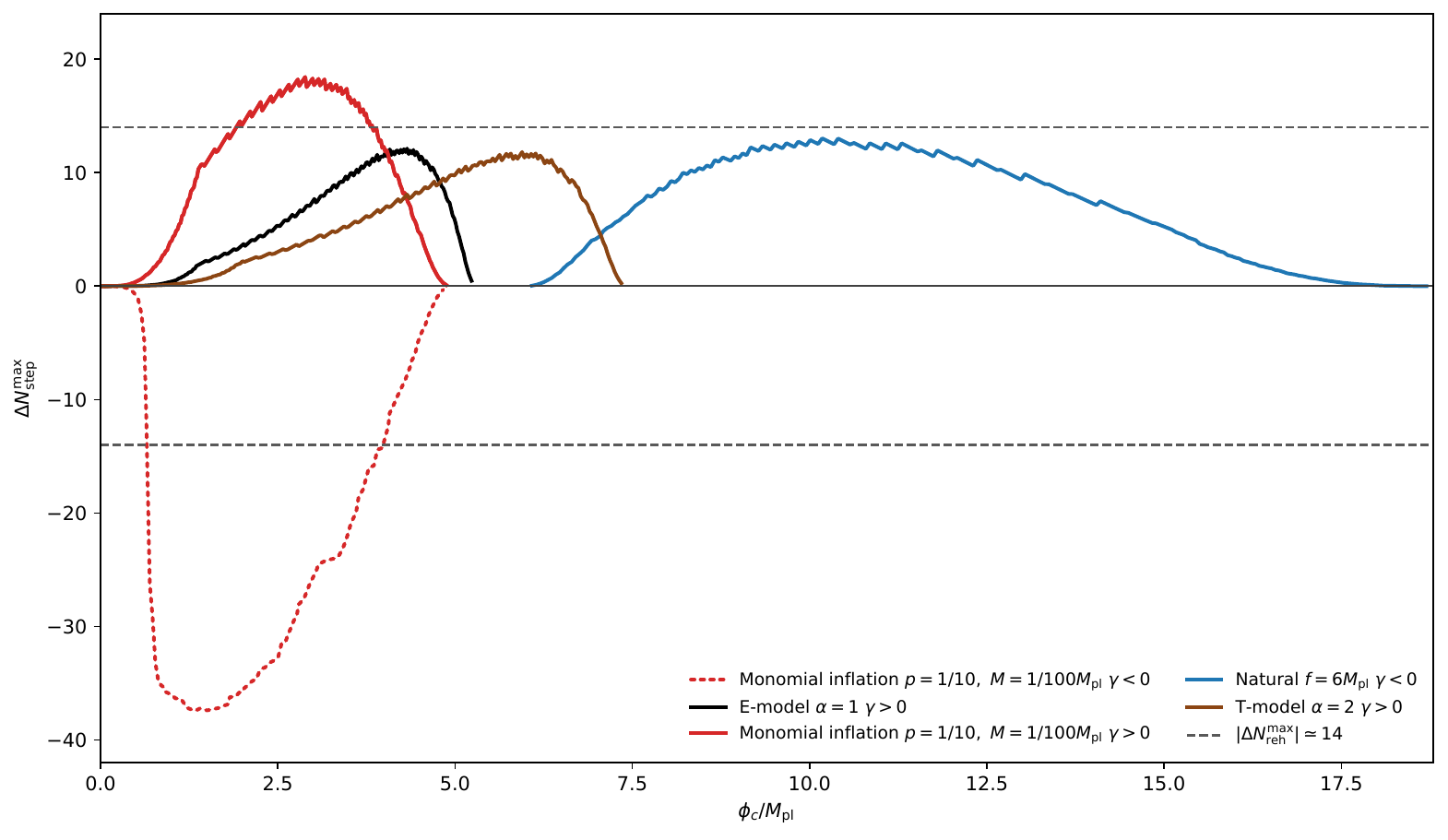}
\caption{\rev{Signed extremal e-fold shift \(\Delta N_{\rm step}^{\max}\) induced by a localized step as a function of \(\phi_c\). For each \(\phi_c\), the plotted value is obtained by scanning over the allowed step parameters. The dashed horizontal lines show the conservative reheating bound \(|\Delta N_{\rm reh}^{\max}|\simeq 14\), and the curves correspond to the benchmark models in Sec.~\ref{sec: Background Dynamics}.}}
\label{fig5}
\end{figure*}

We define the step-induced shift by the effective e-fold displacement along the undeformed slow-roll trajectory,
\begin{equation}
\Delta N_{\rm step}
\equiv
\frac{1}{M_{\mathrm{pl}}^2}
\int_{\phi_{\rm e}}^{\phi_*}
\frac{V_0}{V^{\prime}_0}\, \mathrm{d}\phi
-
N_* .
\end{equation}
 Here \(\phi_*\) is determined from Eq.~\eqref{eq:A1} for the step-deformed potential, with \(N_*=60\) used in the numerical comparison. Positive values indicate that the step shifts \(\phi_*\) farther from \(\phi_{\rm e}\) along the undeformed trajectory, corresponding to earlier horizon exit, whereas negative values indicate the opposite. With the convention of Eq.~\eqref{eq:Delta_N}, this quantity is related to the compensation shift  by \(\Delta N_{\rm step}=-\Delta N_{\rm comp}\). This definition allows a direct comparison between the step-induced remapping and the reheating-induced shift.

For each value of $\phi_c$, we scan over $\gamma$ and $\Delta\phi$, subject to the controlled-deviation conditions in Eq.~\eqref{eq:step_sr_cond} and the localization requirement in Eq.~\eqref{eq:CMB_far_step}. In practice, we implement the latter by bounding the fixed-field deviations of the CMB-scale slow-roll observables,
\begin{equation}
\Delta X(\phi)
\equiv
X(\phi) - X_0(\phi),
\qquad
X\in{n_s,r,\alpha_s},
\label{eq:Delta_X_def}
\end{equation}
where $X$ and $X_0$ are computed from the step-modulated potential $V$ and the undeformed potential $V_0$, respectively. The observables $n_s$ and $r$ are computed from Eq.~\eqref{eq:nsr}, and the running $\alpha_s \equiv d n_s/d\ln k$ is evaluated at leading order,
\begin{equation}
\alpha_s
\simeq
16\epsilon_V\eta_V
- 24\epsilon_V^2
- 2\zeta_V^2,
\qquad
\zeta_V^2
\equiv
M_{\rm pl}^4
\frac{V'V'''}{V^2}.
\end{equation}
We then impose
\begin{equation}
\max_{\phi\in\mathcal{I}_{\rm CMB}}
\left\{
|\Delta n_s(\phi)|,\,
|\Delta r(\phi)|,\,
|\Delta\alpha_s(\phi)|
\right\}
< 10^{-4}.
\label{eq:CMB_nsr_tol}
\end{equation}
This tolerance ensures that the resulting shift is attributable to the background remapping of $\phi_\ast$, rather than to residual direct effects of the step on CMB-scale slow-roll observables.

For the flattening branch of the monomial model, the effective logarithmic slope \(V_0'/V_0+\xi'/\xi\) can become very small. We therefore impose
\begin{equation}
\left|
\frac{V'}{V}
\right|
>
\kappa
\left|
\frac{V_0'}{V_0}
\right|,
\end{equation}
with \(\kappa=10^{-2}\) in the conservative scan. This cut removes the parameter-sensitive near-cancellation regime, yielding a robust estimate of the localized-step remapping.

After imposing these conditions, we define \(\Delta N_{\rm step}^{\max}\) as the signed extremum of \(\Delta N_{\rm step}\) at fixed \(\phi_c\). Specifically, we take the largest positive value on branches with \(\Delta N_{\rm step}>0\), and the most negative value on branches with \(\Delta N_{\rm step}<0\). The resulting signed envelope is shown in Fig.~\ref{fig5}.

The shift in $N_{*}$ induced by post-inflationary reheating can be written as
~\cite{Drewes:2017fmn}
\begin{equation}
\Delta N_{\rm reh}
\equiv
\frac{3w_{\rm reh}-1}
{12(1+w_{\rm reh})}
\ln
\left(
\frac{\rho_{\rm end}}{\rho_{\rm reh}}
\right),
\label{eq:DeltaNreh}
\end{equation}
where $w_{\rm reh} $ is the effective equation-of-state parameter during
reheating. A matter-like reheating phase with $w_{\rm reh}<1/3$ decreases $N_*$, whereas a stiff reheating phase with $w_{\rm reh}>1/3$ increases 
it. Here $\rho_{\rm end}$ denotes the energy density at the end of inflation and 
\begin{equation}
\rho_{\rm reh}
=
\frac{\pi^2}{30}
g_{\rm reh}
T_{\rm reh}^4,
\label{eq:rho_reh}
\end{equation}
with $T_{\rm reh}$ the reheating temperature and $g_{\rm reh}$ the effective number of relativistic degrees of freedom. 

To estimate the largest possible reheating uncertainty, 
we take the conservative values~\cite{Fu:2025ciy} 
\begin{equation}
T_{\rm reh}=4\,{\rm MeV},
\qquad
g_{\rm reh}=10.75,
\qquad
\rho_{\rm end}^{1/4}=10^{16}\,{\rm GeV},
\label{eq0}
\end{equation}
and restrict the equation of state to the causal range $0\leq w_{\rm reh}\leq 1$. This gives
\begin{equation}
\Delta N_{\rm reh}^{\rm max}
\simeq
14.
\label{eq}
\end{equation}
We use this value to define a conservative reheating envelope, shown by the two dashed horizontal lines in Fig.~\ref{fig5}.

Fig.~\ref{fig5} compares $\Delta N_{\rm step}$ with the conservative reheating envelope defined above. The monomial branches can exceed $|\Delta N_{\rm reh}^{\max}|$, especially in the flattening branch with $\gamma<0$. For the plateau-like attractor models and natural inflation, the shifts are typically smaller than this bound but remain comparable to the reheating reference scale. The step-induced remapping is therefore a reheating-scale effect: it can be comparable to, and in some cases larger than, the shift generated by post-inflationary expansion uncertainty, without changing the post-inflationary history itself.

\section{Conclusion}
\label{sec: conclusion}
In this paper, we have shown that a localized step in the inflaton potential can shift the inflationary prediction by locally modifying the inflaton rolling rate and thereby changing the mapping between $\phi_\ast$ and $N_\ast$, as quantified in Eqs.~\eqref{eq:delta_N} and \eqref{eq:compensation_condition}--\eqref{eq:Delta_N}. This remapping shifts the prediction in the $(n_s,r)$ plane in a model-dependent manner and can alleviate the tension between single-field slow-roll inflation and recent ACT data.

At fixed $N_*=60$, the localized step shifts monomial and plateau-like models in the phenomenologically favored directions, substantially improving their agreement with the combined Planck--ACT--LB--BK18 constraints. \rev{For plateau-like attractor models, a positive step moves the pivot point toward the flatter region of the potential, leading to larger \(n_s\) and smaller \(r\). By contrast, in natural inflation the attainable shift remains too small to reach the ACT-favored region without spoiling localization or affecting CMB-scale dynamics.}

\rev{We have also derived semi-analytical expressions for the step-induced remapping using a local-slope approximation in the transition region. The resulting predictions reproduce the numerical results for the representative models and clarify how the step location, width, and height control the displacement of \(\phi_*\) and the shift in $(n_s,r)$ plane.}

\rev{Finally, by defining an effective step-induced e-fold shift $\Delta N_{\rm step}$, we have compared the remapping effect with the uncertainty induced by reheating. We found that the localized step can generate an e-fold shift comparable to, and in some cases larger than, the conservative reheating scale. This provides a minimal phenomenological route to easing the ACT-induced tension in single-field slow-roll inflation: the step acts primarily through background remapping, rather than through direct CMB-scale feature effects or a nonstandard post-inflationary history.}

\begin{acknowledgments}
This work is supported in part by the National Natural
Science Foundation of China under Grants No. 12475067,
No. 12305057, and No. 12235019.
\end{acknowledgments}

\bibliography{reference}  

\end{document}